\documentclass[11pt,a4paper]{article}
\usepackage[latin1]{inputenc}
\usepackage{color}
\usepackage{amsmath}
\usepackage{amsfonts}
\usepackage{amssymb}
\usepackage{graphicx}
\begin{document}
\noindent

\begin{center}

{\fontsize{16pt}{2pt} \bf A biotic cosmos demystified?}

\bigskip
\noindent
\normalsize R.J.Spivey\footnote{Email: yrannwn@gmail.com}\\
\smallskip
\end{center}
\rm

\bigskip
\bigskip
\bigskip
\bigskip
\bigskip

\noindent

\large
\noindent
Oceanic planets formed by type Ia supernovae become spectacularly abundant as stars cease to shine. However, the timing may not be altogether inappropriate. Neutrino annihilation might thermally regulate iron-cored water-worlds, sustaining habitable subglacial oceans. If dark matter and dark energy decay to neutrinos, the universe could support life for $\sim10^{23}$ years. Civilisations surmounting the arduous process of hereditary genetics soon discern the biotic nature of the cosmos and accept their role within it. An infrastructure guards against the spread of rogue colonists. Recruited colonists could harness the available energy for the benefit of life with stupendous efficiency, providing unmistakeable evidence of cosmological optimisation. The anthropic coincidences, inhospitable aspects of the current universe and Fermi's paradox would all be illuminated. Semiconductors sensitive to a neutrinoelectric effect offer a laboratory test of the planetary heating mechanism.
\normalsize

\vspace{400pt}

\begin{center}
\end{center}

\newpage
\setlength{\textwidth}{17cm}
\setlength{\hoffset}{-20mm}

\twocolumn

\noindent
{\large \bf Introduction}

\noindent
Enormous strides have been made towards a physical understanding of the universe, but since the study of the cosmos is as broad a subject as one can imagine, there may be more to comprehending life and its cosmological context than physics alone. Few now dispute the assertion that the conditions for life to emerge are very restrictive. As Freeman Dyson put it ``the universe in some sense must have known we were coming". But if the cosmos was expecting us, it is most mysterious that much of it appears rather hostile to life. In physics, it is the fashion to downplay the significance of the `anthropic' coincidences, and lean instead towards fanciful speculations concerning alternative universes or unobservable regions where nature's laws are randomly arranged and life rarely emerges. Being untestable, such speculations do nothing to bring us nearer the truth. Parallels are apparent with some interpretations of quantum mechanics and the surprisingly pervasive multiverse concept. The popularity and dubious scientific merit of these concepts is useful in highlighting a prevailing bias, suggesting there is some philosophical negligence here that should be guarded against. The mystique of alternative realities carries the risk of detachment from this one and detracts from the goals pertinent to a wider society.

A conflict exists between the presence of anthropic coincidences and the generally inhospitable nature of the universe. The universe does not appear to be catering very effectively to the welfare of its inhabitants. Few planets are habitable and those that are intercept only a minute fraction of the energy radiated by their host star. However, there is at least one way in which this conflict might be resolved. In the distant future, the universe may prove far less hostile to life than has hitherto been appreciated. We may happen to inhabit the universe as it undergoes an initialisation phase and life scrabbles to find its bearings. The proposal essentially identifies this initialisation period with the stelliferous age while stars remain active, after which the targeted annihilation of dark matter within iron-cored oceanic planets sustains life. Direct planetary heating is an extremely efficient means of delivering life-sustaining energy to ice-encrusted oceans. Thus, comfortable aquatic habitats can potentially be maintained for a considerable period of time. Comparing the possible duration of this biotic era to a human lifespan, the stelliferous age might only last one second and the universe may currently be less than a millisecond in age. This proposal predicts interactions testable in laboratories today.

Since this concerns the widespread presence of life throughout the cosmos, we are reminded of another important tension. Recognising that stars are abundant, colonisation guards against catastrophes and extraterrestrials are neither signalling to us nor already here, Enrico Fermi famously asked ``where is everybody?". Hundreds of exoplanets have recently been discovered, and a quarter of solar systems have planets of similar mass to ours orbiting close to their stars \cite{how}. Yet after decades of searching with increasingly sophisticated technology, Fermi's paradox has only deepened. Not one other civilisation has been detected. The maxim `absence of evidence is not evidence of absence' seems apt, but if advanced life is relatively common, why would extraterrestrial civilisations independently, yet unanimously, observe a strict policy of silence? An unenlightening explanation for the paradox is simply that life is rarer than we once thought, despite habitable exoplanets, long-lived stars and a myriad of biophillic characteristics \cite{bar}, \cite{crt}, \cite{crr}, \cite{ree}, \cite{les}, \cite{dav}, \cite{liv}. Few, if indeed any, of these coincidences can be easily construed as being purely anthropic.

Before the discovery of the Hubble expansion, some thought that the universe might be statically maintained by repulsion due to a positive cosmological constant. This was Einstein's well known motivation for not simply assuming $\Lambda=0$. Once a correlation between galactic redshift and distance was established, it became clear that the universe was expanding with time. Although this possibility was encompassed by a known solution of general relativity, the Steady State Theory was later proposed by Hoyle, Gold and Bondi whereby the dilution of matter due to expansion was balanced by the spontaneous creation of matter in empty space. Optimism concerning the long-term future of the cosmos largely evaporated when the microwave legacy of the big bang was first observed in 1965. Though the general view of cosmology has been bleak ever since, there may be a glint of hope. Perhaps it was not the initial optimism that was na\"ive, but the subsequent pessimism.

\bigskip\noindent
{\large \bf A Biotic Cosmos}

\noindent
A model shall now be described hypothesising that the laws of nature and its resources were customised, and indeed optimised in some ways, to support life throughout the cosmos. This notion may not seem tenable at first because it does not correspond to our existing impressions either of the present universe or the future one in which stars will have ceased to shine. However, in a {\it biotic cosmos}, life is sustained for far longer than the time-scales we usually consider. Furthermore, the optimisation has consequences that are measurable physically and appreciable from a strategic perspective. The welfare of life is paramount in this model, physical characteristics of the universe stemming from this requirement. The notion is abandoned that physics is absolute and life is incidental to it. However radical this may seem, this is a manifestly scientific theory making falsifiable predictions.

The biotic era essentially consists of two phases. The first corresponds to the stelliferous age during which intelligent life must be cultivated and highly advanced, trustworthy colonists must be recruited. Beyond the stelliferous age, dark matter annihilation efficiently delivers energy to oceanic planets. Overcrowding can be avoided in these habitats by restricting the rate of energy delivery to the extent that ocean surfaces freeze over, reducing radiative losses at the surface considerably, while the bulk of the ocean remains liquid. For a finite energy budget, this has the effect of greatly extending the duration of the biotic era. With skilled colonisation, life becomes widespread throughout the cosmos and all individuals enjoy a pleasant existence. One sense in which the biotic cosmos model is optimised involves the extremely efficient usage of available energy by conscious lifeforms, this being evaluated over the biotic era's history.

Though no serious discussion of how nature might have been tailored to achieve this optimisation need be attempted here, a mental model is useful when evaluating any optimisation strategy so it may help to outline a simple possibility. We might suppose that benevolent and highly advanced inhabitants of a pre-existing universe found a way of crafting a new universe. Perhaps constrained by a budget of zero energy, they could have configured physics to best suit the needs of life. Furthermore, the physical configuration could have been used to provide a strategic infrastructure safeguarding the long-term welfare of conscious lifeforms against the influence of unenlightened species. As is immediately apparent, such species will inevitably be present at times in an initially lifeless biotic cosmos. Let us further suppose that the configuration process was every bit as painstaking as if these cosmic architects were planning a universe they themselves would be prepared to inhabit. Alternatives to this picture leaving the optimisation strategy unchanged are comfortably accommodated by the model.

An important concept of the biotic cosmos is that conscious lifeforms fall into one of two categories. The term `autons' shall be used to refer to lifeforms who have surmounted hereditary genetics and gained a command of their own biology after undergoing a relatively brief programme of iterative genetic enhancements and refinements. The term is intended to reflect their self-arranged biology, their astronautic propensity, their ethically responsible nature, their self-reliance and independence of governance. Conscious lifeforms still reliant on hereditary genetics and following Darwinian evolution shall be referred to as `zoa'. Where a distinction with zoa may be helpful, technological civilisations who are not autons shall be referred to as `technozoa'. 

Readily apparent improvements in life quality possible using molecular genetics spur technozoa to seize control of their biology. Intellectual faculties and psychological traits, being largely genetically determined, would be amongst the characteristics standing to benefit. Any autons initially ignorant of biotic cosmology are hence unlikely to take long to discover it and comprehend the cosmological context of their existence. Vital prerequisites of colonists are that they possess a firm command of genetics and are ethically responsible. Thus, autons are synonymous with colonists. Biotic cosmology informs them that, with their assistance, high quality life can be widespread in their galaxy for a vast period of time following the stelliferous age. Appreciating that their actions are integral to the success of the biotic cosmos they inhabit, they dedicate themselves to the goal of ensuring that their galaxy will be skilfully colonised once oceanic planets become abundant. In so doing, autons allow themselves to be recruited as colonists cosmologically entrusted with the duty of maximising the welfare of lifeforms inhabiting the post-stelliferous era. Technozoa may or may not appreciate the true nature of the cosmos, that biologically adept colonists are required to colonise the very abundant oceanic planets of the post-stelliferous era or that their planet serves as a nursery attempting to recruit suitable colonists and cultivating innovative genetic solutions to survival challenges. 

In a biotic cosmos, galaxies provide three main lines of strategic defence against inadvertent colonisation by technozoa: (i) the formidable challenge of interstellar travel (ii) tidal-locking of otherwise habitable planets orbiting long-lived stars and (iii) the need for genetic expertise when colonising relatively anoxic subglacial oceans. 

Interstellar travel is extremely challenging, especially during the early stelliferous age when technozoa are most likely to be present. As will be discussed later, even autons may struggle with interstellar travel at this time since interstellar propulsion may be confined to technologies that do not harness the energy of dark matter collected en route, requiring instead that hefty payloads of fuel be taken. Potentially habitable planets orbiting the longest-lived stars (low luminosity red dwarves) rapidly become tidally-locked. The time for the tidal-lock is proportional to the sixth power of the orbital radius \cite{gla}. This also guards against the possibility that technozoa launch life-bearing probes to colonise remote planets. Oceanic planets may offer long-term habitats of anoxic subglacial oceans during the stelliferous age, but these are only attractive to autons capable of formulating lifeforms for these environments. It has been noted that ``nearly all metabolic life-styles will be denied" within Europa's subglacial ocean \cite{gai}. The presence of a liquid ocean beneath Europa's icy surface was suggested some four decades ago \cite{lei} and macrofauna may exist there \cite{chy}.

In the unlikely event that technozoa find means of surviving throughout the stelliferous age without progressing to become autons, a biotic cosmos offers a further precaution. Let us suppose that technozoa establish a galaxy-wide programme of colonisation, suppressing all subsequent recruitment of colonists. Although the probability of this happening may be very low, $p_r\ll 1$, it is unlikely to be zero due for example to tyrannical political regimes or civilisations who have latched onto the notion that genetic enhancement is unethical due to mystical considerations. This can be guarded against this using galaxy mergers. Technozoa would be ousted upon the introduction of auton civilisations in their midst. After $n$ galaxy mergers, the probability of rogue colonists still being present declines as $p_r^n$. Thus, the strategy is an extremely effective one. 

The merger of gravitationally bound galaxies proceeds due to the drag of the inter-galactic medium, whose total mass currently exceeds that contained in galaxies. These mergers serve another purpose -- the formation of a single `super-galaxy'. This is useful since it consolidates the dark matter haloes of $\sim100$ or more galaxies, forming a denser halo better suited to delivering heat to planets throughout the post-stelliferous era. Super-galaxies are anticipated to have fully formed after some $10^{12}$ years, well before the end of the stelliferous age. In order for the strategy of using galaxy mergers to depose rogue colonists to be highly effective, it is necessary for galaxies to recruit colonists prior to the mergers. This suggests that colonists will normally be recruited in most galaxies during the first $10^{10}$--$10^{11}$ years. Since this seems plausible, the prospect of technozoa surviving beyond the stelliferous age might be virtually eliminated.

Survival is an almost constant struggle for most zoa. Evolution breeds lifeforms that reproductively exploit ecological niches. Hence, zoa have a genetic constitution that may be adequate for reproduction but is inadequate for their long-term welfare and survival. Unnecessary hardship can be avoided by ensuring that conscious life does not arise too frequently. This requirement is compatible with the Rare Earth Hypothesis, a model in which the development of complex life is only expected to occur if a highly favourable set of conditions are fulfilled. Although it is commonly thought that simple life may be very common, our understanding of abiogenesis is too limited to be confident of this. There is certainly scope for abiogenesis to be highly improbable in order to ensure that zoa are galactically rare. In addition to restricting opportunities for life to arise, a biotic cosmos imposes sensible time limits on the survival of zoa which fail to progress to become autons.  Few planets remain habitable for much longer than about ten billion years. Due to changes in chemical composition, a star's luminosity varies as stellar evolution proceeds, shifting the circumstellar habitable zone. Rotating planets eventually become tidally locked, depriving them of a defensive magnetosphere leading to atmospheric ablation and inhospitable temperature contrasts. Life-cultivating stars have relatively short lifetimes and once they exhaust their nuclear fuel a pulsating red giant phase follows, purging any life present on orbiting planets. Although this need not involve complete engulfment of a planet by a star, it is thought this will be the ultimate fate of the Earth in several billion years time \cite{sch}.  

In summary, the primary features of the biotic cosmos model are the following:

\begin{itemize}
\setlength{\itemsep}{2pt}
\setlength{\parskip}{0pt}
\item Cultivation of zoa and the rapid nurturing of intelligence via evolutionary stimuli.
\item Avoidance of unnecessary hardship experienced by zoa during the early universe: life is rare and few planets remain habitable for much longer than $\sim10$ billion years.
\item Recruitment of autons during the stelliferous age, either inspired by the inference of biotic cosmology or surreptitiously, through offering enticements to technozoa who surmount hereditary genetics.
\item Provision of a strategic infrastructure guarding against the accidental infestation of a galaxy by technozoa: interstellar travel is very challenging, tidally locked planets are unattractive destinations, anoxic subglacial oceans are difficult to colonise without a thorough command of genetics.
\item Galactic mergers to oust any dominant technozoa suppressing auton recruitment.
\item The generation of large numbers of oceanic planets via type Ia supernovae.
\item The consolidation of dark matter haloes via galaxy mergers for the purpose of interstellar travel once oceanic planets have become spectacularly abundant.
\item Sustainment of aquatic life via the annihilation of dark matter within iron-cored planets with deep surface oceans.
\item Avoidance of overcrowding by restricting the rate of planetary energy delivery.
\item A biotic era supporting life for a period greatly exceeding the stelliferous age.
\item Highly efficient usage of available energy by lifeforms during the biotic era.
\item A framework for elucidating both the Fermi paradox and the Zoo Hypothesis.
\end{itemize}

\bigskip\noindent
{\large \bf Dark Matter - Distribution}

\noindent
In a biotic cosmos, dark matter must play a crucial role, being the primary source of energy that sustains life throughout the biotic era. Baryogenesis has produced a universe with a baryon-antibaryon asymmetry. In contrast, typical dark matter candidates consist of a balanced mixture of matter and anti-matter. Notably, dark matter candidates have not yet had time to self-annihilate. The presence of dark matter is by now very well established in several ways, having long ago been inferred from the motions of galaxy clusters \cite{zwi}. The baryonic dark matter due to MACHOs does not exceed 8\% \cite{tis}. 

The baryon asymmetry, of crucial importance to life, is thought to be due to the CP violation of the weak interaction. Although even small adjustments in the weak force can lead to universes where life is prohibited, the possibility that a universe completely lacking weak interactions might support life has been proposed \cite{har}. By adjusting the other forces and boosting the primordial abundance of deuterium, nucelosynthesis could still proceed, though stellar lifetimes would be curtailed. Within the framework of the biotic cosmos, this hints that flexibility exists to optimise the weak interaction in order to accomplish something which is favourable to life, but in ways that have not been appreciated before. This is plausible if the main benefits of the weak interaction pertain to lifeforms of the future universe. Neutrinos, thought to interact exclusively via the weak force, have played at most an obscure role in the emergence of life. This is most curious given that other particles of the standard model are so clearly of vital importance to life. If neutrinos do play an important life supporting role, what might it be and does it relate to the future universe?

Neutrinos are the only significant component of non-baryonic dark matter identified so far, and indeed the only standard model particles that can contribute to dark matter. Neutrinos and their anti-particles may be equally abundant -- something which is automatically satisfied if they are Majorana particles. Dark matter may be composed of multiple elements, and an appreciable contribution from neutrinos is possible \cite{gaw}. Neutrinos may or may not be the main component of dark matter at the present time, but if they are not, the situation could change radically in the distant future due to particle decay processes. The Super-Kamiokande and SNO experiments \cite{fu1,fu2,fu3} demonstrate that neutrinos undergo flavour oscillations that account for the former anomaly in the observed number of solar neutrinos and firmly establish that they possess mass. Relic neutrinos formed at the big bang compose the hitherto undetected cosmic neutrino background \cite{rin}. Their mass has not been accurately determined but is expected to lie in the range 0.1--2.0 eV/c$^2$. There may be more than three generations and sterile species that lack weak interaction and potentially comprise the cold dark matter (CDM) \cite{ham}. 

Inflationary theories posit that when inflation ended, the inflaton decayed into matter, most of which is non-baryonic. The accelerating cosmological expansion is unlikely to be due to a cosmological constant $\Lambda$ \cite{ldy} and is probably a transient phenomenon \cite{bbm}, \cite{ala}. Many models suggest that dark energy is a scalar field rolling slowly towards zero and coupling to matter \cite{mot}. If dark energy shares features of the inflationary scalar field, it may decay similarly, augmenting dark matter. The possibility exists of a scalar field masquerading as dark matter and decaying into neutrinos \cite{bjd}, or controlling neutrino mass \cite{bro}. Given that neutrinos have such low mass and are very stable against decay, it is plausible that other forms of dark matter may decay in time to neutrino pairs, whether directly ($\chi\chi\rightarrow\nu\bar\nu$) or via metastable intermediate particles. This is not an unreasonable expectation given that of the three generations of particles in the standard model, all ordinary matter belongs to the lightest generation, compatible with stability being most attainable for lighter particles.

Being fermions, neutrinos obey the exclusion principle and are resistant to clumping. Formations of degenerate neutrinos can be modelled by polytropes of index 3/2, assuming spherical symmetry and sub-relativistic momenta \cite{lew}. Polytropes can also represent more general models of dark matter \cite{sax}. Their pressure follows $P=K\rho^{5/3}$ and their central density is proportional to the square of the mass of the neutrino halo $\rho_c\propto M_h^2$. There is no analytical expression for the density variation but it can be computed numerically and is relatively constant over central regions, apparent for example in the figures of Chan \& Chu \cite{cha}. The radius of the formation is well-defined and {\it decreases} as $M_h$ increases according to $R\propto1/\sqrt[3]M_h$. It is interesting that the neutrino mass is sufficiently small that degenerate neutrinos can form haloes larger than galaxies, even though their total mass may greatly exceed that of the contained galaxy.

Above a mass $M_{max}\sim5\times10^{16}M_\odot$, a neutrino cloud will collapse to form a black hole \cite{lew}. This mass is at least ten times that of the $\sim10^{14}$--$10^{15}M_\odot$ currently contained by galaxies gravitationally bound in clusters. This justifies the use of an $n=3/2$ sub-relativistic polytropic neutrino model. Since such galaxies are already bound, a large fraction of them may merge in the distant future, assisted by drag from the inter-galactic medium. The outcome would be super-galaxies with much denser neutrino haloes than those surrounding present day galaxies, particularly if dark matter is almost entirely composed of neutrinos by then. The neutrino haloes of these super-galaxies would be comfortably larger than the stellar disks, so that planets within these galaxies would be surrounded (and permeated) by neutrino dark matter of approximately constant density. Given this scenario, it is interesting to know whether neutrino annihilation could deposit sufficient heat within planets to provide habitation opportunities beyond the stelliferous age. The possibility is far more energy-efficient than stars radiating their heat almost entirely into space. Despite their inefficiency at delivering heat to planets, stars do not use more than 1\% (our Sun, 0.1\%) of their rest-mass energy during their entire lifetime. Thus, the energy squandered by stellar fusion is certainly less than 0.05\% of the total energy of the universe. This is compatible with the possibility that the stelliferous age may only be a brief stage of initialisation in the cosmological history of life.

\bigskip\noindent
{\large \bf Dark Matter - Annihilation}

\noindent
An Earth mass planet's subglacial ocean can be sustained at a habitable temperature by the annihilation of about half a gram of dark matter per second. The discovery that neutrinos have mass established that their physics lies beyond the standard model. As a result, the nature of neutrinos remains unsettled, and different theories have been proposed. The determination of the mass of the various neutrino flavours is ongoing, the number of species has yet to be pinned down, ambiguities concerning chirality and sterile forms await clarification, it is not known whether their mass is Dirac or Majorana and relic neutrinos, second in cosmological abundance only to photons, have eluded detection efforts - prohibiting their physics from being probed in the low energy regime. Given that neutrinos have a mass beyond the standard model, they may also possess electromagnetic properties allowing interactions via virtual photons \cite{tex}. Since it would not benefit the present analysis to overcomplicate the picture, and all that is required is a preliminary assessment of planetary heating adequate to illustrate a lengthy extension of the biotic era, sufficiently robust estimates for neutrino interactions shall be obtained by applying basic concepts of quantum theory. 

\begin{figure}[h]
\begin{center}
\includegraphics[scale=0.8]{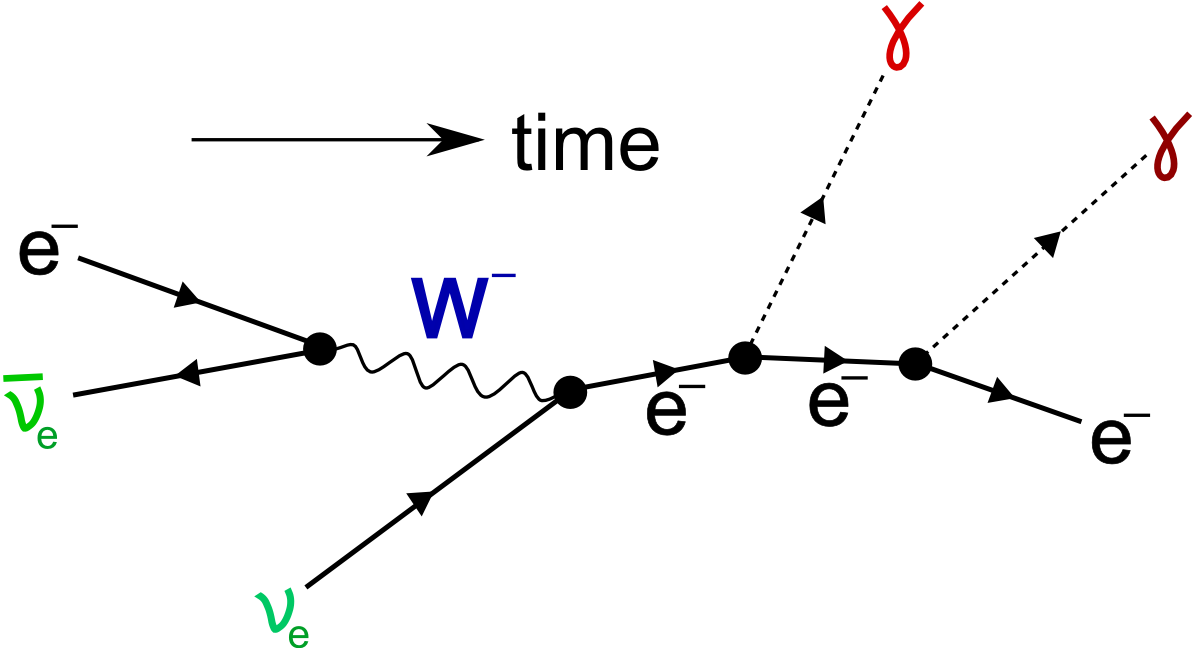} 
\caption{\it Inverse photoneutrino process. A virtual W boson couples to an electron and an antineutrino. Finding itself immersed within a halo of degenerate neutrinos, conditions are favourable for annihilation with another neutrino. An electron emerges in an excited state, decaying here via the emission of a pair of infra-red photons which heat the surrounding matter.}
\label{ann}
\end{center}

\end{figure}

The annihilation of neutrinos via the weak interaction is not anticipated in the absence of other matter, and a spectral trace corresponding to relic neutrino annihilation has not been observed astronomically. The density of electrons in matter greatly exceeds that of neutrinos within the densest haloes. Charged current interactions triggered by the collision of an electron anti-neutrino $\bar{\nu_e}$ and an electron $e^-$ are thus favoured over neutral current interactions triggered by collisions between neutrinos. It is proposed that the reaction $(e^-\bar{\nu_e})\nu_e\rightarrow W^-\nu_e\rightarrow e^-\rightarrow e^-\gamma\gamma$, is the primary mode of relic neutrino annihilation. The radiative emission of an even number of infra-red photons can restore the excited electron to its original ground state atomic orbital, though non-radiative emission of phonons is another possible route for the excited electron to shed energy. Either way, surrounding matter is heated. Figure \ref{ann} depicts the process. Since the laws of quantum physics are reversible in time \cite{fey}, this reaction shall be referred to as the inverse photoneutrino process (IPP), being the time-reversal of the well-established and studied photoneutrino process \cite{rit}, \cite{chi}, \cite{dut}.

Despite the name, weak interactions are not intrinsically feeble, but rather severely limited in range due to the large masses of the W and Z gauge bosons. Electromagnetism and the weak interaction are believed to be manifestations of a unified electroweak theory. Whereas electromagnetism is mediated by massless photons, the W boson has a mass $m_w=80.4$ GeV/c$^2$. A consequence of the Dirac equation is that particles cannot be localised on scales smaller than the Compton wavelength. The reduced Compton wavelength of the W boson is $\lambda_w=\hbar/m_wc\sim2.4\times10^{-18}$ m. Roughly speaking, when an electron $e^-$ and an electron anti-neutrino $\bar{\nu_e}$ are separated by $\lambda_w$ or less, a virtual W boson can couple to them both, triggering interaction. By the uncertainty principle, the W boson can briefly borrow energy for a time $\Delta t<h/4\pi m_wc^2\sim4\times10^{-27}$ s. Providing the relative velocity between the particles is sub-relativistic, this period is not further curtailed, avoiding suppression of the IPP reaction. Since neither electrons bound to atoms nor relic neutrinos are relativistic, the effects of Lorentz contraction can be ignored. Clearly, $\Delta t$ and $\lambda_w$ prohibit the W boson from travelling far. For IPP to proceed, an electron neutrino $\nu_e$ must therefore be immediately available. The width of a real W boson is $\Gamma=2.08$ GeV, corresponding to a mean lifetime of $\hbar/\Gamma=3\times10^{-25}$ s or $\sim75\Delta t$, not enormously longer than the lifetime of a virtual W boson.

In a dark matter halo of neutrinos, degeneracy ensures the de Broglie wavelengths $\lambda_{dB}=h/p=h/m_\nu v_\nu$ of the neutrinos overlap one another. This length is significant because it sets the quantum scale for matter waves that arise due to wave-particle duality and are associated with diffraction and wavefunction solutions to the Schr\"odinger equation. Max Born successfully interpreted the wavefunction as a probability amplitude whose squared magnitude represents the probability of a particle being present. Hence, the de Broglie wavelength reflects the uncertainty concerning a particle's location or, as diffraction experiments demonstrate, the range of a particle's sensitivity to other particles. Thus, when $e^-\bar{\nu_e}\rightarrow W^-$ occurs within degenerate neutrinos, conditions are favourable for IPP to proceed. An electron neutrino can sense the presence of a W boson within its matter wave, offering a tantalising opportunity to annihilate.

The virtual W boson acts as a mediator for IPP but in a sense, so too does the (real) electron. The necessity of an electron being present prevents the wasteful annihilation of neutrinos in empty space. Once the electron and the anti-neutrino are within the coupling range of a virtual W boson, all that is required for $\nu_e\bar{\nu_e}$ annihilation to proceed is for a $\nu_e$ to be able to make itself available to the W boson. Whereas scattering between particles can be thought of as being caused by a force that involves the exchange of many force-mediating bosons, the exchange of a single W boson is sufficient for mediating particle annihilations. In the limit where the anti-neutrino has a small energy compared to the electron's ($E_\nu\ll E_e$), the cross-section for $e^-\bar{\nu_e}$ scattering can be obtained as \cite{tex},\cite{hoo}:

\begin{equation}
\nonumber
\frac{d\sigma}{dE_e}\approx\frac{G_F^2E_em_e(9E_e+5m_e)}{8\pi E_\nu^2}
\end{equation}

This expression illustrates how the scattering cross-section grows as the neutrino's energy decreases. One expects an equivalent term to be present in the denominator of an expression for the neutrino annihilation cross-section at low energies. Physically, this corresponds to improved opportunities for interaction via the uncertainty principle as the relative velocity between electrons and neutrinos declines. This behaviour is unlikely to extend to extremely low energies since a plateau will be reached once the relative velocity is sufficiently low that the dynamical time-scale associated with the potential is comparable to the crossing time. However, other enhancements also exist at low energies. While perturbative expansions are normally employed in the calculation of scattering and annihilation cross-sections, it is only possible to ignore higher order terms for relativistic particles. Perturbative approximations break down when the kinetic energy is not large compared to the potential energy, and plane-waves no longer adequately represent the physics. In general, such considerations can have very significant effects on cross sections of particles with low relative velocities. For example, in the context of Sommerfeld corrections during WIMP annihilations, a $1/v_{rel}$ S wave enhancement and a $1/v_{rel}^3$ P wave enhancement is expected, with possible resonances at specific energies \cite{ien}. In a biotic cosmos, one would not be surprised to find that the low energy annihilation of neutrinos preferentially occurs in specific circumstances that assist the sustainment of life.

\bigskip\noindent
{\large \bf Targeted Regulation}

\noindent
At room temperature, the vast majority of electrons bound to atoms occupy their ground-state, and thus have no surplus energy to donate. Since neutrinos have mass, there is a minimum temperature which has to be exceeded before Maxwell-Boltzmann statistics allow significant numbers of electrons to be excited to energies above the threshold needed for neutrino pair ($\nu\bar{\nu}$) production. IPP will proceed at a rate that is largely insensitive to the ambient temperature. However, $\nu\bar{\nu}$ production will suddenly become significant above some threshold temperature, offering a means of cooling which becomes exponentially more effective with temperature as the fraction of electrons with the threshold energy increases. In equilibrium, IPP is counteracted by $\nu\bar{\nu}$ production and the extraction of thermal energy flowing through the planet before radiating into space. This equilibrium regulates both the maximum internal temperature of a planet and its surface heat flux.

\begin{figure}[h]
\begin{center}
\includegraphics[scale=0.5]{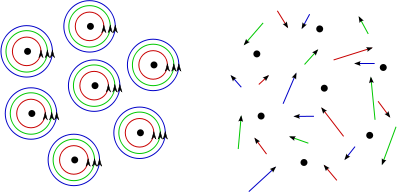} 
\caption{\it Opportunities for IPP require the presence of electrons but whereas $\nu\bar{\nu}$ production can cool degenerate matter (right) to very low temperatures, its energy threshold prevents the excessive cooling of non-degenerate matter (left). In atoms, electrons occupy the lowest available energies and have little or no surplus energy. In cold electron degenerate matter, electrons move freely and generally have kinetic energies well above the threshold for $\nu\bar{\nu}$ production.}
\label{ato}
\end{center}
\end{figure}

Although this method of planetary heat delivery is far more efficient than stellar radiation, its inefficiency might rise sharply if it operated also within stars, brown dwarves and gas giants. Fortunately, this is rather unlikely. These bodies are so large and their internal pressures so high that their electrons become degenerate and delocalised from atoms. Beyond the stelliferous age, all stars will become electron degenerate white, black or brown dwarves, the same being true also of gas giants. Electrons bound to atoms would only be present in surface regions. Degenerate electrons are capable of cooling matter through $\nu\bar{\nu}$ production, even at arbitrarily low temperatures. Terrestrial planets differ from all these in being composed of atoms whose electrons are bound to them and, at moderate temperatures, have energies near the ground state meaning they have insufficient surplus energy for $\nu\bar{\nu}$ production. Since electrons have approximately a million times the mass of neutrinos, their typical kinetic energies in degenerate stars are far higher than the energy required for $\nu\bar{\nu}$ production. Being so far above the threshold, IPP can do little to resist their cooling. Thus, very little energy is wasted on them. Figure \ref{ato} depicts this simple but effective means of discriminating between planets requiring warmth and other objects which should be allowed to cool.

It follows that in gas giants, stars, white dwarves and black dwarves, which will account for the majority of all baryonic matter, the balance between IPP and $\nu\bar{\nu}$ production is likely to be almost exact, conserving the neutrino dark matter. In contrast, planets are receptive to energy delivery via IPP if they are sufficiently cool. Lacking accurate knowledge of the $\nu_e$ mass, and the rate at which $\nu\bar{\nu}$ production occurs relative to other processes such as photoemission, no firm estimate as to the regulation temperature shall be attempted here. However, the anticipated range for the neutrino mass suggests that the core temperature will be several thousands of Kelvin. An energy scale of 1eV corresponds to a temperature of 11,600$^\circ$K, but only a small fraction of electrons need have a surplus energy above the $\nu\bar{\nu}$ production threshold during thermal regulation. Less than 1 in $10^{24}$ electrons need participate in IPP each second, but the participation fraction will typically be well above this and almost fully counterbalanced by $\nu\bar{\nu}$ production. The Boltzmann distribution controls the population levels of electrons at various energies, establishing thermal regulation.

It is necessary to avoid neutrino annihilation in the surface regions of an oceanic planet because regulation only operates at a certain temperature well above that of either surface ice or the water immediately beneath it. It will often be the case that neutrino heating has much in reserve, and regulation is necessary to prevent overheating in surface regions where the temperature remains below the regulation point due to rapid cooling from the surface nearby. This can be prevented if neutrino annihilation proceeds exclusively within a planet's iron core, being suppressed elsewhere. This is possible if the neutrino annihilation energy corresponds to energy transitions of bound electrons in hcp iron at high pressure but not to. Since IPP occurs only in the presence of electrons, and electron energies are restricted by quantum mechanics, IPP only occurs when conditions are favourable. Relic neutrinos have energies comparable to the coarse energy transitions present in iron. Their low temperature makes them highly monoenergetic with kinetic energies a few millionths of their rest mass energy. This means that neutrino energies are unlikely to closely match the energy differences between the quantum states available to electrons bound to atoms in vacuum or other low pressure environments. Being a transition metal with a rich electronic structure and many similar energy levels within its outermost 3{\it d}/4{\it s}/4{\it p} shells, iron is very well suited to IPP involving neutrinos whose mass are tuned accordingly. Within a planet's core, iron is present in the hcp phase, its density is significantly increased and the energy levels of the electrons undergo pressure broadening due to the Stark effect. These considerations may also be effective in blocking the delivery of heat to bodies rich in iron but smaller than planets, and to surface regions of gas giants where hydrogen is present in the atomic and molecular state.
 
\bigskip\noindent
{\large \bf Oceanic Planets}

\noindent
Supernovae are highly efficient at synthesising heavy elements, especially the common type Ia variety. Observations have determined that, by mass, their ejecta consist of 18\% oxygen, 15\% silicon, 13\% iron, and 49\% nickel, almost all in the unstable form $^{56}$Ni, along with small amounts of carbon, calcium, sulfur and magnesium \cite{tan}. These emerge in layers, the outermost layer of oxygen emerging with highest velocity and the ferromagnetic elements with least velocity. Oxygen reacts with hydrogen as it ploughs through the interstellar medium (ISM), forming water which later cools to ice. Obviously, water is of tremendous importance to the biochemistry of lifeforms, their foodstuffs and their habitats.

The progenitors of these supernovae are carbon/oxygen white dwarves of a consistent mass near the Chandrasekhar limit of $M_{Ch}\approx 1.4M_\odot$. They are likely to be explained by a single explosion scenario \cite{maz}, and triggered by accretion of matter from a binary companion. Stars must exceed $M_{Ch}$ if they are to generate supernovae other than the type Ia variety. At present, only a small fraction of stars are so massive \cite{kro}. Star formation will continue well into the future, but the initial mass function will be increasingly skewed towards lower mass stars \cite{dok}. In terms of long-term planet formation, type Ia supernovae will utterly dominate the process.

The majority of stars have at least one companion, lone stars like the Sun are in the minority. Some stars in binary systems may have originally been planetary gas giants orbiting a white dwarf. Accretion of gas from the ISM and IGM proceeds rapidly for objects with sufficient gravity to retain an atmosphere of light elements. Gas giants orbiting close to a white dwarf accrete gas very rapidly, far more so than the compact stars they are orbiting. Since only a small fraction of baryonic matter is luminous, given sufficient time, binary star systems can accrete enough matter from the ISM that a type Ia supernova occurs. Thus, a substantial fraction of baryonic matter will eventually undergo nucleosynthesis in type Ia supernovae, each of these generating enough material to form about half a million Earth-mass oceanic planets.

Nature has the capacity to assemble planets of a consistent size from supernova ejecta. Water can form ice XI, a ferroelectric phase of ice relevant to planet formation \cite{ied}, \cite{fuk}. The agglomeration of ice XI fragments proceeds by long-range electrostatic attraction \cite{wah}, forming comets. Forces between magnetic dipoles strongly dominate gravitational forces over ranges below $\sim17$ km according to an inverse fourth power law. Ferromagnetic atoms align themselves to an imposed magnetic field to form linear chains rather than ring formations \cite{huc}. Linear chains will clump together via short-range dipole attraction. Magnetic agglomeration is thermally regulated by loss of magnetic ordering above Curie temperatures, these being $631^\circ$K, $1403^\circ$K and $1043^\circ$K for Ni , Co and Fe respectively. Long range forces between the ferromagnetic elements $^{56}$Ni and $^{56}$Co are further inhibited by energy released during radioactivity decay to $^{56}$Fe, half lives being 6 days for $^{56}$Ni and 70 days for $^{56}$Co. Large objects cool very slowly, so protoplanets also have the capability to regulate their long-term growth. 

Having higher velocities than protoplanets, comets have large, elliptical orbits. As planets cool, comets can slowly deposit sufficient water to form oceans several tens of km in depth. Over the course of time, these oceanic planets become extremely abundant, accounting for a significant fraction of all baryonic matter. It has been estimated that oceanic planets are already quite common \cite{ehr}. It is both significant and important that upon freezing, water forms a low density solid. Since ice floats on water, it can serve to thermally insulate a body of water. Habitable subglacial oceans can be maintained when heat is present within the core of a planet.

\bigskip\noindent
{\large \bf Planetary Heating}

\noindent
If an oceanic planet's core is maintained at a well-defined temperature, the heat flux through the planet's surface is also well-defined, almost independently of planet size. Planets are of a size providing sufficient gravity to ensure sphericity. Hence, after tectonic activity has abated, their oceans assume a uniform depth. If planets were much larger then the ratio of their volume to their surface area would increase: more material would be required for a given volume of inhabitable ocean. There is evidence here that baryonic matter has been efficiently used and that oceans are cosmologically important. Also, we notice that terrestrial planets have a surface gravity that permits the development of intelligent fauna lacking gross blood pressure variations caused by their vertical scale.

The density of electrons bound to atoms within a planet is around $n_e\sim3\times10^{30}$ m$^{-3}$. Less than two electrons per atom are used for metallic bonds in transition elements such as iron \cite{fle}. A cylinder of radius $\lambda_W$ and length $l$ contains $\pi l\lambda_W^2n_e$ electrons. If the electrons are stationary, the mean free path between collisions triggering IPP will be $1/\pi\lambda_W^2n_e$. But, since the electrons move with velocity $v_e=\alpha cZ/n\sim0.06c$, and a velocity relative to the neutrinos of $v_{rel}=\sqrt{v_e^2+v_\nu^2}$, the neutrino velocity being $v_\nu$, the mean free path will be $v_\nu/\pi v_{rel}\lambda_w^2n_e$. The merger of a cluster of galaxies with a mass at the lower end of the scale could form a neutrino halo of $M_h=10^{14}M_\odot$, radius $R_h=5.6\times10^5$ ly and central neutrino density $\rho_c=1.9\times10^{-21}$ kg/m$^3$.

It is assumed that neutrinos have a mass in the range 0.2--1 eV. The characteristic neutrino velocity at a temperature of $2^\circ$K would be $v_\nu=\sqrt{kT/m_\nu}=0.011c\sim0.027c$. The velocity of the virtual W boson arising during IPP will resemble that of the seed electron since its mass greatly exceeds the neutrino's. The W boson finds itself bathed in degenerate neutrinos, which due to their relative motion, have a de Broglie wavelength of $\lambda_{dB}=h/m_\nu v_{rel}=100\sim220  \mu$m. The typical separation between neutrinos would be $\sim\sqrt[3]{m_\nu/\rho_c}=5.7\sim9.8 \mu$m, so conditions are favourable for IPP throughout the neutrino mass range. In this example, the mean free path for neutrinos between IPP opportunities lies within the range 2.7--5.9 km. A surface heat flux of $0.1$ W/m$^2$ demands that an Earth-sized planet be delivered heat at a rate of $\sim4.7\times10^{13}$ W. Here, the total mass of neutrinos within a planetary volume is $\sim2$ kg, and it is sufficient that about half a gram annihilate per second. This translates into a requirement that, on average, neutrinos annihilate within a planet after travelling less than $4000v_\nu\sim14$ million km.

Since the mean separation between IPP opportunities is less than a millionth of this, neutrinos can very comfortably deliver the necessary heat in a regulated fashion. This conclusion remains intact even if a suppression factor of 75 representing the ratio of the mean lifetime of a real W boson relative to a virtual W boson is allowed for. During regulation, an equilibrium will be established between neutrino annihilation, neutrino production and heat delivery, providing a controlled heat flux to planetary oceans. This example highlights the plausibility of this mechanism for a low mass super-galaxy, and its ability to continue functioning even as the mass of the neutrino halo becomes depleted by two or three orders of magnitude, enough to encompass the current Milky Way $\sim3\times10^{12}M_\odot$ even before its anticipated merger with Andromeda several billion years hence. It can be seen that the mechanism is relatively insensitive to the neutrino mass, allowing the mass to be set so that (i) a dark matter halo is large enough to contain all anticipated galaxy sizes and (ii) haloes have insufficient mass to collapse gravitationally and (iii) the neutrino annihilation energy is not dangerous to living tissue and (iv) it may assist interstellar travel at later times.
 
\bigskip\noindent
{\large \bf The Biotic Era}

\noindent
Life can be supported for considerably longer than the stelliferous age if neutrino dark matter selectively annihilates within oceanic planets. The duration of the biotic era depends on the abundance of habitable planets and the rate at which heat is delivered to them. For a fixed habitation volume, the mean space available to each living organism is inversely related to the rate at which energy is delivered. However, if the rate of energy delivery falls below some threshold, life cannot be supported at all: subglacial oceans have the potential to freeze completely. A biotic cosmos may exploit neutrino annihilation to sustain planets at habitable temperatures with overcrowding being avoided, or even minimized, by restricting heat delivery. Though not a goal as such, a consequence of this could be a lengthy biotic era. In this section, a general upper estimate for the duration of the biotic era will be obtained independently of any specific model of dark matter or annihilation mechanism, assuming only that annihilation occurs almost exclusively within iron-cored planets of similar size to ours.
\begin{figure}[h]
\begin{center}
\includegraphics[scale=0.19]{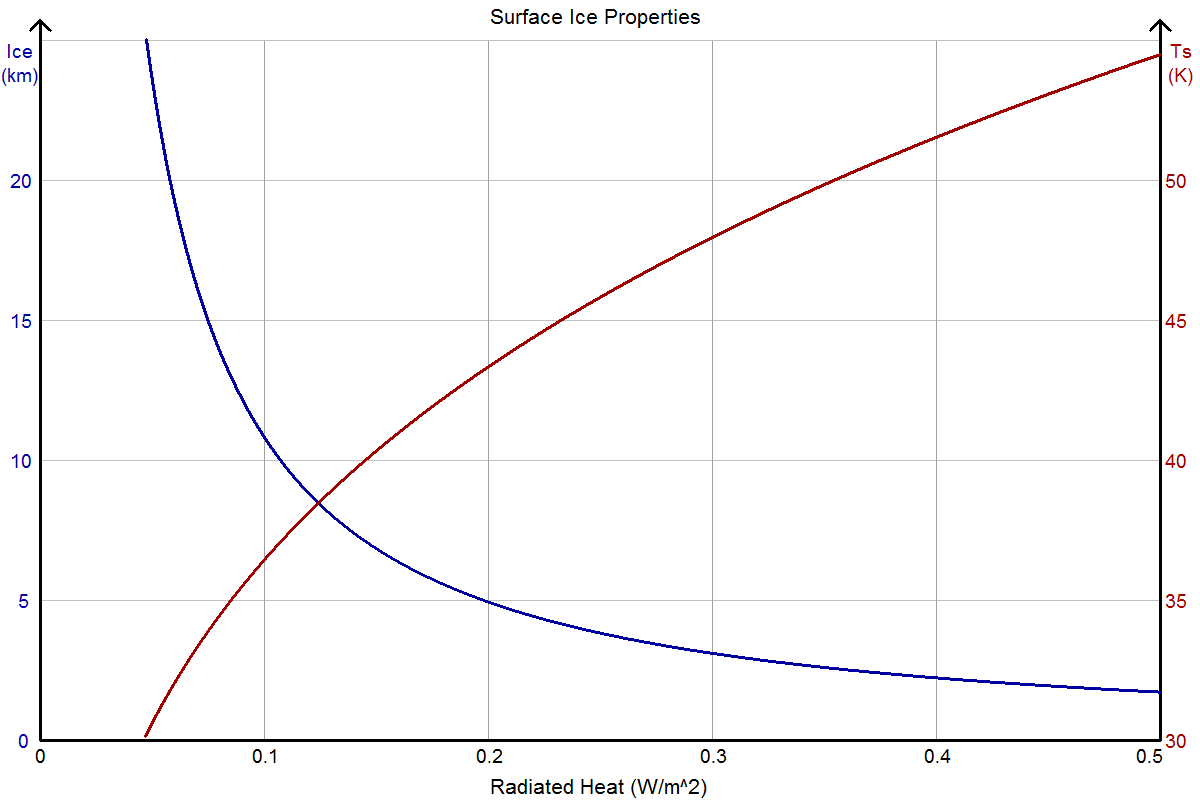} 
\caption{\it Ice thickness (blue trace) and surface temperature (red trace) plotted against heat flux at the surface of a dark matter heated subglacial oceanic planet with a rarefied atmosphere.}
\label{ice}
\end{center}
\end{figure}

The sublimation pressure of ice at cryogenic temperatures is negligible \cite{fei}, important in preventing the long-term loss of water from a planet. Planets of the post-stelliferous age are therefore assumed to have negligible atmosphere and heat loss will proceed via black body radiation from the surface given by $4\pi r^2\epsilon\sigma T_s^4$. Here, the surface temperature is $T_s$, the planet radius is $r$, the surface emissivity is $\epsilon$ and $\sigma$ is the Stefan-Boltzmann constant. The surface temperature decreases as the thickness of the insulating surface ice increases. Approximately speaking, living space is maximised when the volume of ice is approximately the same as the volume of the subglacial ocean for planets with shallow oceans.

Allowing for some differences between planet sizes and compositions, a combined ice and ocean depth of 20km or more shall be assumed, a conservative estimate given the 18\% oxygen composition of the ejecta of type Ia supernovae and the anticipated consistency in planet masses. The thermal conductivity of hexagonal ice Ih is inversely related to temperature but is relatively constant with pressure. It shall be approximated here by $\kappa=546/T$ in units of W/m.K. The surface emissivity for black body radiation is taken to be unity. No significant convection is anticipated within surface ice at relevant thickness. At a heat flux $Q$, the ice thickness is $h_{ice}\approx 546\ln(265/T_s)/Q$ where $Q=\sigma T_s^4$. If $T_s=36^\circ$K then $Q=0.1$W/m$^2$ and the required ice thickness is $h_{ice}\approx$ 11 km. The relationship between $Q$, $h_{ice}$ and $T_s$ is plotted in figure \ref{ice}, illustrating the compromises involved as the living space is maximised. It can be seen that $Q=0.1$W/m$^2$ is a reasonable choice to within a factor of two or so. For the same heat flux, the surface ice could be reduced somewhat if the surface emissivity were decreased due to contaminants often present in comets. Volcanic activity might produce a CO2 atmosphere when the planet was newly formed, freezing in a thin surface layer as the planet cooled. Solid CO2 has a thermal conductivity approximately one sixth that of water ice at the same temperature, but the surface layer would be too thin to have much impact on the optimal value of $Q$.

The boiling point of water increases with pressure so the temperature of the deepest regions of a subglacial ocean could be several hundreds of degrees Celsius. Superheated water has low viscosity and is remarkably effective at decomposing many substances, useful if there is biotic waste to be reprocessed \cite{bru}. At pressures equivalent to a few km in depth, there is no abrupt change in the density of water above the critical point T$_c$=374$^\circ$C and temperatures even beyond T$_c$ are in principle tolerable. At constant temperature, the density of water increases with pressure. But, if the temperature also increases with pressure, the density can decrease. If the mean density decreases with depth, strong hydrothermal circulation occurs. A convective ocean has the advantage of reducing stagnation and assisting nutrient transportation. The redistribution of heat by thermals and B\'enard cells reduces the mean vertical temperature gradient. For the density of water to decrease with depth, $h$, it is necessary that d$T/$d$P\gtrsim0.8^\circ$K/MPa or equivalently d$T/$d$h\gtrsim0.007^\circ$K/m.
\begin{figure}[h]
\begin{center}
\includegraphics[scale=0.47]{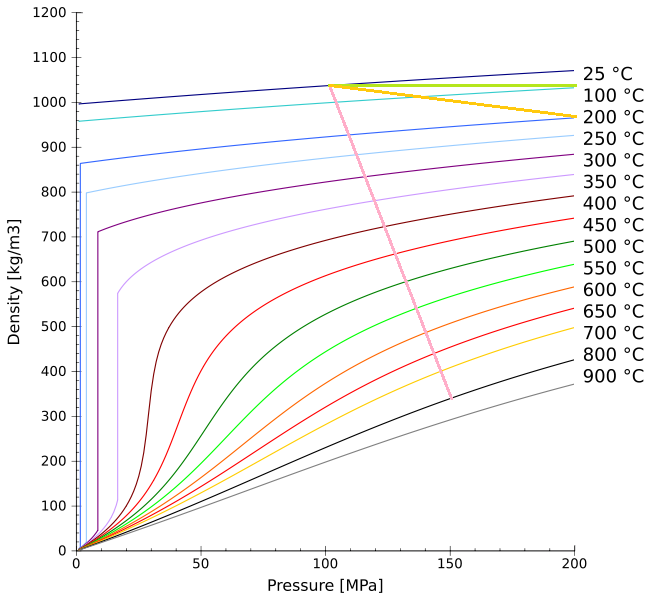} 
\caption{\it Variation of water density with pressure for selected temperatures. A pressure of 200MPa corresponds to a depth of approximately 20km beneath the upper ice surface. Convection requires that the density decreases with depth. Straight lines depict pure conduction (pink), moderately (amber) and highly effective convection (green) for a heat flux of 0.1W/m$^2$.}
\label{wtr}
\end{center}
\end{figure}

The vertical temperature gradient, assuming Earth gravity, is only $\sim0.2^\circ$K/km for an adiabatic ocean. Although the heat flux is small, it invalidates an assumption of adiabaticity. Were it not for convection, a heat flux of $Q=0.1$W/m$^2$ would correspond to a strong gradient of $\sim250^\circ$K/km, represented by the straight pink line in figure \ref{wtr}. However, the Rayleigh number is of order $10^{17}-10^{20}$ indicating a potentially convective ocean where chaotic thermals permit the gradient to be substantially reduced. The straight amber line in figure \ref{wtr} illustrates the estimated effect of this convection. Even if this convection is highly effective, a temperature gradient of at least $5^\circ$K/km appears necessary, otherwise convection ceases altogether as depicted by the straight green line in figure \ref{wtr}. However, the full 3D situation is more complicated. Qualitatively, hot and cold spots will be present in the oceans e.g. due to hot plumes in the mantle and variations in the thickness of the sea bed and underlying mantle. Vertical streams of water will redistribute heat in fast moving water columns, driving a complex mixing process tending to equalise the overall thermal contrast within the ocean. These coarse estimates suggest that a 20km deep ocean might be half frozen, an acceptable fraction. Though the ocean thermodynamics are extremely complex, their influence only improves the quality of life by increasing the habitable volume of the liquid ocean and providing a useful temperature gradient. Planets with very large water fractions could have layers of ice VI, VII and perhaps also ice X beneath their liquid oceans, forming solid-state convective mantles \cite{fu}. However, heat transport through the oceans was apparently not modelled in that analysis, accounting for the surprisingly consistent ocean bottom temperature of 283K. These high pressure phases of ice (2.2GPa$\lesssim P_{VII}\lesssim$60GPa) prohibit oceans deeper than $\sim100$ km for Earth-like planets.

Only about 12\% of baryonic matter is luminous at present. In all, baryonic matter makes up some 5\% of the total mass-energy of the universe ($\Omega_B\sim0.05$). Non-baryonic dark matter comprises around 22\% of the universe ($\Omega_{DM}\sim0.22$) and the enigmatic dark energy associated with the surprising discovery that the expansion of the universe is accelerating accounts for virtually all of the remaining 73\% ($\Omega_{DE}\sim0.73$). Although few planets have formed by now, planet formation will proceed for as long as stellar evolution continues and type Ia supernovae occur. If the final fraction of baryons having been processed by type Ia supernovae is $\xi$, and we assume that few of the resulting planets collide with stars, then oceanic planets will comprise $\Omega_P\sim\xi\Omega_B$ of the universe. Because binary systems are common, and 90\% of stars have a mass below $M_{Ch}$, it is anticipated that $0.01\lesssim\xi\lesssim 0.40$. Since $\Omega_B\ll 1$, it is not crucial that $\xi$ be large. 

To gauge an upper limit for the duration of the biotic era, it is now assumed that a net annihilation of dark matter occurs only within planets and that dark energy decays in such a way as to augment dark matter. The energy available to sustain life on each planet cannot exceed $(\Omega_{DM}+\Omega_{DE})/\Omega_P\approx 19/\xi$. It is assumed that each planet has a mass matching that of the Earth, a radius $r=6100$km and dissipates heat at a rate of $4\pi r^2Q\approx 5\times10^{13}$ W. For $\xi=0.02$, the duration of the biotic era is then $\sim3.5\times 10^{23}$ years and for $\xi=0.2$, its duration is $\sim3.5\times 10^{22}$ years. Note that these are briefer than the time-scales of proton decay (at least $10^{34}$ years) and electron decay (at least $6.4\times 10^{24}$ years). Overall, a good ballpark estimate for the duration of the biotic era would therefore be $10^{23}$ years.

\bigskip\noindent
{\large \bf Dark Matter Conservation}

\noindent
If this is a biotic cosmos, the present phase of accelerating expansion must somehow serve to benefit life. A galaxy cluster is a gravitationally bound object with an acceptable $10^{14}$--$10^{15}M_\odot$ mass for a habitable super-galaxy. However, an upper limit exists to the mass of a neutrino halo $M_{h_{max}}\sim5\times 10^{16}M_\odot$ beyond which degeneracy pressure cannot resist collapse \cite{lew}. The merger of multiple clusters could exceed this limit but dark energy driven expansion may be effective in preventing this. The product of a merger between all galaxies in a cluster would possess a neutrino halo whose radius is $\sim8/m_\nu M_{h_{max}}^{1/3}$ \cite{lew} or $2.6\times10^5$ ly, enough to contain a large galaxy. The central density would be $\sim 3.2\times 10^{-20}$ kg/m$^3$, six times the mean halo density. A degenerate neutrino halo was recently inferred from the lensing data for the large Abell 1689 galaxy cluster \cite{nie}.

If neutrinos are astrobiologically important, their capture by black holes is of concern. This is exacerbated by the tendency of degenerate neutrinos to repopulate any voids or regions of low density. Unless black hole growth can be halted, rapid neutrino depletion can occur for black holes above a critical size. Supermassive black holes at the centres of most galaxies contain $\sim0.2\%$ of the mass of the galactic bulge \cite{fer}. However, no correlation, inverse or otherwise, with dark matter is yet perceptible \cite{kor}, suggesting that these black holes are at present too small to have a significant impact on dark matter haloes. The rate of dark matter depletion is controlled by the surface area of a black hole originally formed by the collapse of other forms of matter. Since $R_{\rm BH}=2GM_{\rm BH}/c^2$, this is slow initially because $A_{\rm BH}\propto M_{\rm BH}^2$. The time taken for a supermassive black hole to fully deplete a neutrino halo can be estimated as:
\begin{equation}
\nonumber
t_{BH}\approx\int_{M_{\rm BH}}^\infty{\frac{dM}{4\pi \sigma_\nu R_{\rm BH}^2}}\approx\frac{c^4}{16\pi G^2\rho_\nu M_{\rm BH}}\sqrt{\frac{m_\nu}{kT_\nu}}
\end{equation}

For a neutrino mass of 0.3eV, a neutrino temperature of $\sqrt[3]{4/11}T_{cmb}=2^\circ$K, a central halo density of $\rho_c=2\times10^{-19}$ kg/m$^3$ corresponding to a total halo mass of $\sim10^{15}M_\odot$ and an initial black hole mass of $4\times10^{9}M_\odot$, the neutrino halo would be exhausted after $t_{BH}\sim10^{17}$ years. Clearly, if we inhabit a biotic cosmos, this is puzzling. Why might there exist an impressively efficient mechanism for sustaining life over a $10^{23}$ year period, only for this to be ruined relatively early on in many galaxies by supermassive black holes formed in the early universe apparently serving no vital purpose? Some potential solutions to this complication are outlined below:

\begin{itemize}
\setlength{\itemsep}{0pt}
\setlength{\parskip}{2pt}
\item{Advanced civilisations must attend to this problem themselves by deploying neutrino extractors in orbit around the central black hole. Though not impossible, this seems to be contrary to the biotic cosmos scenario.}
\item{The growth of black holes is strongly suppressed above some threshold in mass e.g. $10^{11}M_\odot$. Since Hawking radiation is minuscule, this would require new physics.}
\item{Black holes differ from classical black holes and somehow repel neutrinos.}
\item{Dark matter is primarily something else.}
\item{Angular momentum produces neutrino formations of zero central density \cite{mad}.}
\item{The decay of dark energy alters the vacuum density, triggering the decay of black holes to a true underlying vacuum and resulting in their contraction to zero size.}
\end{itemize}

Though theoretically abstruse, the latter possibility is appealing. It is likely that the mass of a black hole is accurately balanced by its negative potential energy. Hawking radiation is associated with a minute black hole temperature, evidence that the total energy is approximately zero. If so, the event horizon may be a boundary prohibiting negative energy voyeurism. It seems plausible that a vacuum phase transition beyond the horizons might permit the decay of black holes in the future. The deflation of a black hole due to phantom dark energy has already been considered \cite{bab}. It may be that black holes can only decay once the underlying vacuum reaches a sufficiently low energy density. If so, it would be natural to expect black holes to start contracting following the decay of dark energy to dark matter. If this is the case, one might infer that the presence of black holes in the universe was not inevitable, suggesting their presence at the centres of galaxies has served some function e.g. seeding structure formation or regulating the growth of galaxies. Since it is so difficult to discern any important role for black holes in a biotic cosmos, one might even speculate that dark energy ultimately decays {\it through} them, resulting in the release of neutrinos precisely where they are most needed: within galaxies. 

\bigskip\noindent
{\large \bf Biotic Energy Efficiency}

\noindent
If life is unique to our planet, never migrates to other solar systems and never arises elsewhere, the biotic efficiency of the universe will be extremely small. The Sun provides a mean energy flux at the Earth's surface of $\sim 342$ W/m$^2$. The energy supplied by the lowest rungs of the food chain, the net primary production, is a small fraction of the available power $\sim 1$ W/m$^2$. The luminosity of the Sun is steadily increasing and the Earth may only be habitable for a further billion years. The total energy consumed by conscious life throughout the history of the solar system is then the product of the net primary production and $4\pi R_E^2\Delta t$, about $3\times 10^{31}$ J. The biotic efficiency can be calculated by relating this to the energy of the observable universe, giving an utterly minute result of $\sim10^{-38}$.

Neutrino dark matter, augmented by the future decay of dark energy, might sustain life within subglacial oceans for $\sim10^{23}$ years. Since the heat deposited by neutrinos within planets necessarily flows through the subglacial oceans, it is accessible to the lifeforms there. The efficiency with which life utilises the available energy would be limited only by the skill of those autons who designed the ecosystems and the inhabitants perhaps continually refining them. 

Since dark matter and dark energy comprise almost all the energy density of the universe ($\Omega_{DM}+\Omega_{DE}\sim 0.95$), conscious lifeforms might potentially harness a large fraction of it, even to the extent that the biotic efficiency approaches unity. The prospects for ultra long-term life in the universe have been considered elsewhere \cite{dys}. Although $10^{23}$ years is a long time, it is not the duration of life that has been maximised in the biotic cosmos model, but the biotic energy efficiency. The length of the biotic era is a reflection of the avoidance of overcrowding, not of eking out an existence for as long as possible. 

Since autons form stable civilisations, they have ample time to study the cosmos thoroughly and refine their physical theories. They will obtain from this tight constraints on the biotic energy efficiency of the cosmos. If ours is a biotic cosmos then autons will be aware of the fact because the biotic energy efficiency is remarkably close to unity, a highly improbable outcome in a randomly organised universe. The most rational explanation for such a result would be that the universe was pre-configured in the full anticipation that colonists would arise and competently discharge their duties. 

\bigskip\noindent
{\large \bf Life in the Milky Way}

\noindent
Biotic cosmology would have a profound influence on the mindsets of autons. Together with their highly rational nature, it is reasonable to expect a strong convergence in their motivations and activities. The psychology associated with their cosmological outlook demands a total reappraisal of our preconceived expectations regarding extraterrestrial life. Before proceeding to consider this, note that it is immediately apparent that in a biotic cosmos there is no need to evaluate various terms in the Drake equation. The model predicts that colonising civilisations will be relatively rare in galaxies at the present time. Large galaxies must recruit colonists during the stelliferous age but in order to avoid unnecessary hardship, zoic populations must be kept low. Though star formation continues for most of this period, it steadily declines and the fraction of newly formed stars capable of cultivating life also decreases due to the trend towards lower mass stars. The ousting of any rogue colonists is assisted if genuine colonists are present when larger galaxies merge to form super-galaxies. Recruitment is therefore expected to succeed during the early stages of the stelliferous age, ($\sim10^{11}$ yrs). Hence, it is unlikely that there are many colonising civilisations present in this galaxy now. Indeed, it is quite plausible that there are none. Due to their highly transitory nature, technozoa will be much rarer still than autons, and the possibility of hearing from them is very remote.

However, we ought not discount the possibility that a few colonising civilisations are already present in this galaxy. In a biotic cosmos, it would be surprising if there were more than, say, ten of these civilisations already present. Unless one happens to be located in close proximity to us, any messages they may broadcast could be undetectable. It is therefore necessary to consider whether these colonists are engaging in long-range interstellar travel, or remain highly localised within the galaxy. Interstellar travel benefits from the en route collection of fuel. Being composed of both matter and anti-matter, it is possible to envisage advanced technologies which harvest relic neutrinos for this purpose, though perhaps not before galaxies have merged. Since the density of a neutrino halo is proportional to the square of its total mass, the present neutrino density within this galaxy may be several orders of magnitude less than its value once galaxies have merged. This may delay long-distance travel until harnessing dark matter for propulsion becomes more practical. That time coincides with oceanic planets becoming highly abundant. Degenerate neutrinos have the useful property of streaming in to fill a void. This could be exploited by streamlined spacecraft trailing a lengthy series of regularly spaced neutrino harvesters in their wake. For now, inferior propulsion strategies might permit the restricted colonisation of small numbers of local solar systems, adequate to guard against extinction by most natural catastrophes. Since the universe is very young, planet formation has barely begun and there is no need yet for widespread colonisation, autons can instead avail themselves of the preparation time to attend to the enormous task they have undertaken.

Curiosity may drive autons to visit other solar systems in order to ascertain the frequency with which abiogenesis occurs, study the progress of evolution on other planets, collect genetic samples from other lifeforms and try out their skills at ecosystem design. Even if interstellar travel is very challenging and costly at present, they may still consider this to be a worthwhile exercise. Although the biotic cosmos model permits the rarity of colonists to be estimated, and requires that conscious zoa be rare to limit the hardship they experience, it does not constrain the rarity of unicellular or multicellular lifeforms lacking consciousness. We remain extremely ignorant of the challenges involved in the transitions involved in developing a biochemistry that supports consciousness. Perhaps autons will share a little of this ignorance. If life is generally very rare, then the curiosity of autons may not be sated by visiting solar systems in their immediate neighbourhood. However, even a million solar systems is but a small fraction of a galaxy. Thus, it is plausible that autons might not have spread far throughout the galaxy yet.

Nevertheless, let us also consider the scenario in which autons are already widespread, having populated a subset of the most desirable planets in the galaxy. They would then be capable of contacting any technozoa in their vicinity should they wish to do so. In order to assess whether they would decide to proceed, we can draw on what the biotic cosmos model allows us to infer about their psychology. There are two important aspects to bear in mind. The first is their awareness of biotic cosmology and the purpose in life that this gives them -- to strive towards galactic colonisation and maximise the overall welfare of lifeforms throughout the biotic era. It may be no exaggeration to say that they perceive this as their raison d'\^etre. Secondly, having carefully refined genetics, as individuals they would be extremely advanced intellectually. They would also lack the eccentricities, idiosyncrasies and foibles which are prevalent in zoa. They would be capable of highly rational thought and arriving at consistent opinions and conclusions, even on relatively complex matters. The moral difference between right and wrong is likely to be perfectly clear to them on many issues we would have no hesitation in describing as a matter of opinion. This consistency would apply across individual autons and independent auton civilisations. Therefore, concerning the issue of whether or not to meddle in the development of zoa, autons are likely to observe a consistent policy.

Given that their life's work is devoted to what they discern to be a cosmologically decreed goal of colonisation, a decision to contact (highly transitory) technozoa would not be undertaken lightly. There are strong reasons to suspect that autons would deem it inappropriate. Autons would not want to encourage the infestation of the galaxy by technozoa, a galaxy they are carefully preparing to colonise. Aware that the recruitment of colonists is a natural process, it would be arrogant of autons to assume they could improve on it, a quality they are most unlikely to possess. As advanced as they are, autons would realise that nature was configured by superior intellects to theirs, capable for example of predicting accurately the conditions required for abiogenesis and the rates at which complex biochemistry develops and evolution proceeds. Any interference in the recruitment of other colonists might seem unthinkable to them. 

They are also likely to see this in terms of minimising the risk of compromising the welfare of vast numbers of future lifeforms: those inhabiting the colonies of the post-stelliferous era. Since nature nurtures zoa at a cosmologically pre-determined rate, autons are likely to respect that and refrain from activities such as seeding planets with zoa, inhibiting the development of naturally cultivated zoa, making technozoa aware that their galaxy already contains colonists or interfering in any other way with their development or recruitment as colonists. Any desire to acquire competent collaborators or suppress competition from other colonists is unacceptable according to inferences drawn from biotic cosmology. Providing zoa with assistance risks detracting from nature's scheme to maximise welfare over the entire history of the biotic era. Colonists would benefit greatly from learning of innovative solutions to the challenges of survival, solutions which may be prohibitively difficult or too time consuming for them to obtain directly. If planets hosting zoa are fairly rare, they will be all the more valuable to autons.

Autons are aware that they inhabit a biotic cosmos from which they derive a common goal and purpose in life. They are naturally guided by biotic cosmology and essentially have two choices concerning zoa -- meddling in their development or allowing the natural process of evolution and colonist recruitment to proceed in a cosmologically ordained fashion so that the overall welfare of lifeforms alive during the biotic era benefits. It is here argued that during the early universe when few planets have formed and colonisation has not commenced in earnest, the latter of these is the most responsible option. Due to their highly rational nature and the presence of safeguards on the duration of zoic life, a unanimous concordance amongst autons on this issue would not be surprising. The following principle, based upon biotic cosmology, may help summarise the proposed stance they might take:

\bigskip
\noindent
{\it
Colonists best serve the needs of a biotic cosmos by allowing nature to play those roles it is configured for and highly effective at while colonists focus on tasks the cosmos is reliant upon them discharging competently and without distraction. }
\medskip

This `duty principle' leaves colonists free of the burden of attending to the needs of zoa. Contact with technozoa would be prohibited since that would amount to meddling with the recruitment process. Therefore, they would be free to focus their efforts on preparations for the colonies of the post-stelliferous era, working with the cosmos to ensure that colossal numbers of sentient lifeforms enjoy a pleasant existence during the biotic era. Technozoa are a small and extremely transient subset of zoa, most of whom are incapable of intelligent dialogue but are of no lesser importance. Nature limits the hardship of both. Being more advanced, technozoa require less assistance than other zoa, not more. The emergence of technozoa on a planet might spark interest amongst autons, but would do nothing to persuade those autons already aware that the planet had hosted zoa for millions of years that their existing policy of non-intervention required revision. Even if extraterrestrials broke with protocol to assist life on Earth, we might see them first attending to the welfare of animals. The duty principle might also have some bearing on the question of the fate of zoa sharing the same planet as newly recruited colonists.

Although autons would be reticent to communicate with us, communication with other auton civilisations would be an altogether different matter. It would be natural for them to form collaborations since they would lack hostile tendencies and the welfare of future lifeforms stands to benefit. When establishing contact with one another, a covert means of communication would be necessary to prevent technological zoa from eavesdropping. This would be relatively straightforward for autons who may reason that detecting pulses of low-energy (1 keV or less) mono-energetic neutrinos is virtually impossible for technozoa. Stealth aside, providing their energies are well above those of relic neutrinos, this approach also offers some advantages over electromagnetic communication channels in terms of signal distortion when passing through the ISM. Since the harvesting of neutrino energy may be crucial to interstellar travel, colonists are likely to develop means for selectively absorbing neutrinos of various energies, making this a highly attractive mode of covert communication. Our civilisation seemingly lacks the know-how to detect low energy neutrinos.

Fermi's paradox has survived a few decades of searching for radio signals. Yet, the limited sensitivity only imposes loose constraints on the number of actively messaging civilisations presently within our galaxy. The ability of the biotic cosmos to provide direct estimates for the rarity of autons supplies far more stringent constraints, predicting that life will eventually be common in virtually every galaxy. Despite this, the biotic cosmos retains compatibility with Fermi's paradox, and thereby even offers a resolution of the paradox, for the following reasons:

\vspace{-3pt}
\begin{itemize}\itemsep0pt
\setlength{\parskip}{2pt}
\item The stelliferous age has barely begun.
\item Statistically, galaxies need not recruit many colonising civilisations.
\item Hardship is lessened if rudimentary conscious lifeforms do not frequently arise.
\item Technological zoa are rare due to their very fleeting and transient nature.
\item Interstellar travel is difficult until galaxies merge, forming dense neutrino haloes.
\item Planet formation delays colonisation.
\item Technozoa represent an infestation risk that autons will be unwilling to assist.
\item Colonist recruitment could be hindered if the presence of autons were apparent.
\item The cosmos is effective at recruiting colonists. Hence, autons consider meddling with this process counter-productive and a distraction from their appointed responsibilities.
\item Independently recruited colonists have the option of using covert communication.
\end{itemize}

Life being rare does not resolve Fermi's paradox: an explanation for that rarity is required. The biotic cosmos supplies such an explanation. In popular fiction at least, advanced life is often portrayed as being common and a contradiction with the lack of real-life evidence for extraterrestrials is sometimes avoided. Olaf Stapledon's 1937 novel Star Maker featured an alien race choosing to conceal their presence from pre-utopian primitives incapable of space travel. The Zoo Hypothesis \cite{bal} is a general model in which extraterrestrials consistently observe some policy of non-intervention towards life here, as exemplified by Star-Trek's prime directive. This consistency may be universal or stem from guidance provided by a well-respected civilisation. Nevertheless, the underlying motivations for such policies require elucidation. The biotic cosmos supplies a framework that can underpin this concept. But, rather than a zoo, a planet harbouring zoa is more accurately regarded as a colonist recruitment nursery and factory of biological innovation. 

In this light, SETI's negative results provide reassurance that if colonists are in our neighbourhood, they are responsible ones. History has repeatedly taught us we are not as special as we previously thought. The world is not flat, the Sun does not revolve around the Earth, many other solar systems have planets, we are not at the centre of the universe, our DNA barely differs from that of other species, our `precision cosmology' is very na\"ive, anthropic coincidences are neither anthropic nor coincidental, extraterrestrial civilisations do not share our priorities or our genetic frailties and, being both ancient and sophisticated, would be unlikely to learn anything from us.

\bigskip\noindent
{\large \bf Experimental Investigation}

\noindent
Do we inhabit a universe optimised for life, or do we instead inhabit one that is some thirty eight orders of magnitude less energy efficient towards life? Since neither of these drastically contrasting possibilities can be ruled out at present, this highlights the enormity of our cosmological ignorance. According to the biotic cosmos model, an early exit awaits lifeforms who fail to appreciate and avail themselves of the opportunities presented to them by nature. This imposes a useful limit to the hardship they inflict on themselves and others. Potential benefits associated with surmounting hereditary genetics are not in doubt, but it would be useful to ascertain more conclusively whether or not neutrino annihilation is astrobiologically important.

If, as the present analysis suggests, neutrino annihilation can maintain subglacial oceans in the post-stelliferous era then regardless of whether or not we inhabit a biotic cosmos, the long-term prospects for life could be radically improved. Either way, experiments that can detect relic neutrinos, infer their mass and investigate their annihilation rate are of great importance. Since neutrino annihilation delivers heat within large volumes of pressurised iron in the presence of a high density neutrino halo, the prospects of directly detecting this heat in laboratory settings seem very slim. Once equilibrium is achieved, planetary heating proceeds at a mean rate anticipated to be $\sim10^{-7}$ W/m$^3$. Out of equilibrium, this rate would only be sufficient to change the temperature by around one degree every million years. The internal or surface temperatures of planets in thermal equilibrium and remote from the Sun could be measured, but would first need to be discovered, and would only yield ambiguous results due to complicating factors such as geothermal flux due to radiogenic heating, dissipation of primordial heat, and the presently low relic neutrino density.

Since neutrino annihilation within oceanic planets is mediated by electrons bound to atoms, with the annihilation energy being transferred to those electrons, opportunities exist to sense the presence of the excited electrons using technologies routinely used in electronics. In particular, semiconductors are materials in which a forbidden energy band separates a valence band from a conduction band. These forbidden bands, or band-gaps, typically span energies comparable to the neutrino annihilation energy, which in the case of relic neutrinos, is dominated by their rest mass. The energy contribution due to their degeneracy will be small, making them highly mono-energetic at present within this galaxy. An excited electron could be promoted across the band-gap from the valence band to the conduction band. From there, it might emit infra-red photons, generate phonons or participate in current flow if a potential gradient is presented. The latter possibility, where electrical current arises due to electron mediated annihilations of neutrino pairs, shall be referred to as a neutrinoelectric effect.

A very sensitive detector is obtained if a material with the right band-gap is operated at low temperature. Band-gaps can be coarsely set by blending elements in chosen ratios. The number of electrons thermally excited into the conduction band will be extremely low at cryogenic temperatures so that the presence of conduction electrons due to neutrino annihilation is expected to cause a very noticeable increase in the conductivity of the semiconductor as the band-gap energy is decreased to match the threshold required for neutrino annihilation. Finding this neutrinoelectric effect requires investigating the energy range corresponding to the possible neutrino mass. The bandgap energy can be modulated or finely tuned e.g. by the application of an electric field, pressure or by varying the temperature. The promotion of an electron into the conduction band is forbidden unless the bandgap is smaller than the neutrino annihilation energy, so a step change in conductivity is expected when the band-gap crosses this threshold.

Since relic neutrinos are so mono-energetic, the transition zone could equate to a bandgap shift of 0.01\%$\sim$0.1\%. The conductivity shift would therefore be very abrupt. The input bias currents of some leading operational amplifiers are measured in tens of femto amps, comfortably low enough to permit the detection of relic neutrino annihilation at the expected rate in a modest quantity of semiconductor. A wafer with a volume of $\sim30$ cm$^3$ might yield a current of 1pA due to the annihilation of relic neutrinos. Assuming a signal is obtained and the knee point determined, a test confirming that relic neutrinos are responsible is possible. This would involve tracking the knee point as a function of time, which should drift slightly due to daily variations caused by planetary rotation and the Earth's orbit around the Sun. 

Semiconductors sensitive to the neutrinoelectric effect have much to offer in the detection of relic neutrinos and the determination of the absolute mass of an electron neutrino. Such an experiment might determine the neutrino mass to a high precision at relatively low cost, or allow refinements of mass measurements obtained from experiments like KATRIN whose resolution is quite poor, especially at the lower end of the mass range. Accurate knowledge of the neutrino mass would be of potential use in the design of devices capable of trapping relic neutrinos and harvesting their energy. There may not be many commercial applications for such devices on Earth at the present time since solar energy is plentiful and easily collected, at least when sunlight is present. However, technologies could be developed that augment present SETI strategies.

\bigskip\noindent
{\large \bf Cosmological Timeline}

\noindent
Beyond the stelliferous age, galaxies evolve in several ways due to gravitational effects. When an interloping star enters the planetary system of another and passes as close to a planet as the star it is orbiting, then the planet is likely to be perturbed sufficiently for it to be ejected from the system. The spatial density of stars within the Milky Way is $\sim0.003$ ly$^{-3}$. The distance between encounters at range $R=1$ AU is $\sim1/0.003\pi R^2\approx5.10^{11}$ ly. If stars typically have relative velocities of $v_s\sim150$km/s, a conservatively high estimate, then planets are ejected from solar systems on a timescale of $\sim10^{15}$ yrs. Thus, few planets are ejected from solar systems during the stelliferous age, allowing stars to serve as signposts to planet locations. Upon ejection, oceanic planets are better described as free-roaming water-worlds.

Dynamical relaxation and gravitational encounters between stars lead to their evaporation from a galaxy on a timescale of approximately $\sqrt{NR_g^3/GM}\sim10^{19}$ years where $R_g$ is the radius of a galaxy of mass $m_g$ containing $N$ stars of mass $M$, with around 90\% to 99\% of stars being ejected \cite{dys}. Inhabited planets can be actively steered e.g. using dark matter to ensure that they remain within the galaxy. This would be necessary in any case to guard against occasional collisions. This leaves inhabited planets with uncontested access to the galaxy's dark matter halo. This may help boost efficiency if uninhabited but water-deficient planets were receiving dark matter heating. 

After some $10^{24}$ years, gravitational radiation causes galactic orbits to decay \cite{dys}. This extremely gradual process takes even longer than the duration of the biotic era. Planetary navigation can easily correct for this with negligible energy loss. Since neutrino haloes would be spherically symmetric, actively steered planets can navigate orbits away from the original galactic plane. This would also reduce the risk of collisions, the only penalty being additional isolation from other colonies. 

The sketch of a biotic cosmos is now complete. Arranging the primary ingredients in chronological order, the cosmological timeline is summarised in the following table:

\begin{center}
\begin{tabular}{|c|c|}
\hline {\bf Cosmological Event} & {\bf Time (yrs)} \\
\hline
\hline {\it Big Bang} & 0\\
\hline {\it First stars} & $\sim10^9$\\
\hline {\it First supernovae} & $\sim10^9$\\ 
\hline {\it First planets} & $\sim10^{9.1}$\\ 
\hline {\it Simple life} & $\sim10^{9.6}$\\ 
\hline {\it Conscious life} & $\sim10^{9.7}$\\ 
\hline {\it Colonists} & $\sim10^{10}$ \\ 
\hline {\it Red Giants} & --- \\ 
\hline {\it Galaxy mergers} & $\sim10^{11}$\\ 
\hline {\it Dark energy decay} & $10^{11}-10^{12}$\\ 
\hline {\it Black hole dissipation} & $10^{11}-10^{12}$\\ 
\hline {\it Abundant planets} & $\sim10^{13}$\\ 
\hline {\it Full colonisation} & $10^{13}-10^{14}$\\ 
\hline {\it Stars become dark} & $\sim10^{14}$\\ 
\hline {\it Free-roaming planets} & $\sim10^{15}$\\ 
\hline {\it Steered planets} & $\sim10^{16}$\\ 
\hline {\it Stellar clear-out} & $\sim10^{19}$\\ 
\hline {\it Biotic era ends} & $\sim10^{23}$\\ 
\hline {\it Collapse} & ???\\ 
\hline
\end{tabular} 
\end{center}

\bigskip\noindent
{\large \bf Discussion}

\noindent
Attempts to identify a uniquely self-consistent grand theory of physics have been faltering for some time, and an acceptance already appears to be emerging that this may be a forlorn hope. While the pursuit of an understanding of nature is admirable, the insistence that nature conforms to personal expectations is questionable. If no unique theory of physics exists, it is interesting to know whether or not physics is arbitrarily configured. Given the biophillic characteristics of the universe, it stretches credulity to believe that physics is arbitrary, and attempts to support this hypothesis which invoke alternative physics in unobservable regions of this universe and others, do nothing to ease this, being entirely reliant on fanciful speculations. The biotic cosmos posits a falsifiable scientific model in which physics is non-arbitrary. Having been optimised for life, physics might not be unique in a biotic cosmos, but it would certainly be very special. Since the welfare of lifeforms is of paramount importance, any position of proud contempt attempting to portray this as a disappointing outcome would be exposed as such and difficult to defend.

Consciousness is something we are all familiar with, but neither physics nor biochemistry allow us to infer that complex organisms possess sentience. It is na\"ive to expect physicists to satisfactorily describe the cosmos using the same physics that has failed to make any tangible progress towards accounting for how lifeforms are able to experience consciousness. It is therefore extremely unlikely that extraterrestrials would would a similar avenue of investigation when pursuing cosmological truths. If this is a biotic cosmos, one expects any mature extraterrestrial civilisation to have discerned the fact and thereby comprehended the cosmological context of their own existence, however that conclusion may have been reached.
 
This model contends that civilisations understanding that they inhabit a biotic cosmos and sufficiently advanced to serve as colonists will appreciate they have a cosmological role to play, and willingly comply. They will thereby be recruited as colonists and dedicate themselves to their duty of working with the cosmos to ensure that colossal numbers of sentient lifeforms enjoy a pleasant existence during the post-stelliferous era. This is only possible if the highly abundant oceanic planets are responsibly colonised. Planets currently cultivating rudimentary life are of no urgent importance to colonists. If such planets succeed in recruiting further colonists, this will only assist the goals of a biotic cosmos. If not, evolution produces highly innovative solutions to the challenges of survival, which colonists would find valuable and may be interested in examining as evolution reaches its natural conclusion. In the meantime, existing colonists best serve the needs of a biotic cosmos by allowing nature to do its work while they focus on their appointed tasks. Any overlap in these activities might be considered counter-productive and therefore best avoided. 

Short of the exploration of deep space or extraterrestrial contact, no previous resolutions of the Fermi paradox have provided testable predictions. In a sense, this proposal is the first scientific resolution of the Fermi paradox. Compelling evidence to support this model is potentially obtainable by studying the neutrinoelectric effect within semiconductors. Should this convincingly establish that the laws of nature are highly optimised for life, then it would be logical to infer the necessity of colonists in each galaxy, and thereby how rare intelligent life is likely to be without any recourse to the Drake equation. We might also appreciate that our planet could be described as a colonist recruitment nursery and that we may be undergoing a form of quarantine because extraterrestrials are not so irresponsible as to meddle with the natural order. In short, the cosmos might no longer seem so mysterious.

It is not impossible for modifications of the dark matter composition or annihilation mechanism to be accommodated within this model, though there would be tight constraints on any alternatives to neutrinos. The primary hallmark of a biotic cosmos is a high biotic efficiency and other forms of dark matter annihilation might in principle be able to achieve similar efficiencies. However, neutrino annihilation also appears optimised in the sense that neutrinos form collapse resistant haloes large enough to engulf entire galaxies so that energy is accessible throughout the contained galaxies. Also, neutrinos are effective in their avoidance of planetary overcrowding due to their very gradual annihilation. When these three factors are considered, it would be surprising if any other dark matter candidate were capable of competing with neutrinos. Other details are yet more difficult to adjust such as the aquatic habitats afforded by the oceanic planets that become so abundant by the end of the stelliferous age or the necessity for colonists to possess considerable genetic expertise in order to exploit these opportunities well. If we inhabit a biotic cosmos then there are profound implications both for the long-term welfare of life and our comprehension of nature. It would not be unreasonable to infer that time transcends the universe and may even be eternal. We would need to abandon the notion that physics is absolute and accept instead that it is reconfigurable, even though we may never know how.

Uncertainty would still shroud important details even if it can be robustly established that neutrino annihilation has the capacity to greatly extend the duration of the biotic era. Might there be other components of dark matter which do not serve any purpose in planetary heating? Will black holes deplete neutrino haloes, bringing things to a premature close? Is the cosmological acceleration set to continue indefinitely or is it a transient and dynamical phenomenon? Is it possible that the decay of dark energy does not trigger the disappearance of black holes? These are undeniably serious concerns but they may all be countered by one fact. An anthropic coincidence that was inferred from the presence of others, had no bearing on human evolution and transforms the long-term prospects for life throughout the universe would be neither anthropic nor coincidental. This could not simply be dismissed on tautological grounds -- a criticism historically levelled at anthropic coincidences. Instead, this would be intriguing and potent evidence that we inhabit a biotic cosmos, providing confidence that life will eventually thrive in each galaxy.

Freeman Dyson dared to defy the pessimistic philosophy espoused by the renowned molecular biologist Jacques Monod, encouraging others to do likewise \cite{dys}. In {\it Chance and Necessity} Monod wrote ``Man at last knows that he is alone in the unfeeling immensity of the universe: neither his destiny not his duty have been written down". Monod considered existence pointless and life an irrelevant accident of nature. Yet, something moved him to share his views with others, making it difficult to take his remarks seriously. A biotic cosmos is far from indifferent to life, and mature extraterrestrial civilisations may be convinced that they inhabit one. If so, Monod's views would seem abhorrent to them. A colonist recruitment process may be guarding against civilisations who do not prioritise the welfare of their successors. For colonists, the next $10^{23}$ years may seem very rosy - whether with us, or without us.

\onecolumn
{}

\vspace{123pt}
\begin{figure}[h]
\begin{center}
\includegraphics[scale=0.23]{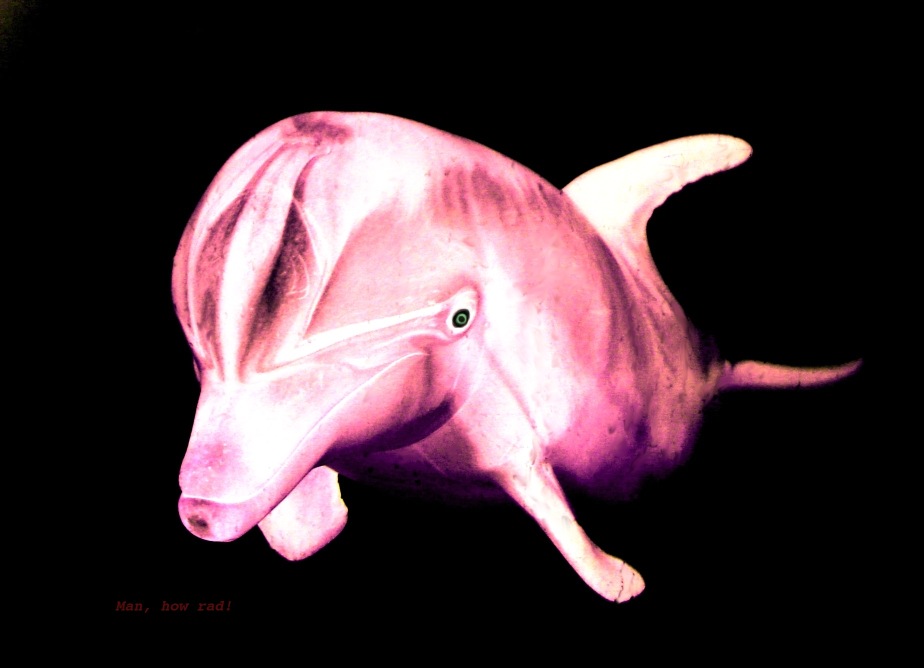}
\label{wat}
\end{center}
\end{figure}

\end{document}